\documentclass[12pt]{article}
\usepackage{amssymb}

\def\nn{ \nonumber }
\def\bq{ \begin{equation} }
\def\eq{ \end{equation} }
\def\ben{ \begin{eqnarray} }
\def\en{ \end{eqnarray} }
\def\frac#1#2{{#1\over #2}}
\def\dfrac#1#2{{\displaystyle{#1\over#2}}}

\newtheorem{prop}{Proposition}
\newtheorem{thrm}{Theorem}

\begin{document}

\title{On the Lax pairs for the generalized Kowalewski and Goryachev-Chaplygin tops}
\author{V. V. Sokolov \\ 
Landau Institute for Theoretical Physics,\\
Kosygina str. 2, Moscow, 117334, Russia \\
\it\small e-mail: sokolov@landau.ac.ru\\\\
A. V. Tsiganov\\
 Department of Mathematical and Computational Physics,\\
 St.Petersburg University,
  St.Petersburg, 198904, Russia\\
\it\small e-mail: tsiganov@mph.phys.spbu.ru
 }
 \date{}
\maketitle

\begin{abstract} {\small A polynomial deformation of the Kowalewski top is considered. 
This deformation includes as a degeneration a new integrable case for the Kirchhoff 
equations found recently by one of the authors. 
A $5\times 5$ matrix Lax pair for the deformed Kowalewski top is proposed. Also deformations of 
the two-field Kowalewski gyrostat and the $so(p,q)$ Kowalewski top are found. 
All our Lax pairs are deformations of the corresponding Lax representations  
found by Reyman and Semenov-{T}ian {S}hansky. In addition, a similar deformation of the 
Goryachev-Chaplygin top and its $3\times 3$ matrix Lax representation is constructed.

}
\end{abstract}

\section{Introduction.}
\setcounter{equation}{0}

In the paper \cite{sok2} it was shown that a top-like system which corresponds to 
the following Hamilton function 
\begin{equation} \label{tailHAM}
H=J_1^2+J_2^2+2 J_3^2
-2(c_2x_1-a_1)J_3-2 a_1 c_2 \,x_1-c_2^2\,x_3^2-2 c_1\, x_2\,,   
\end{equation}
where $c_1,c_2$ and $a_1$ are arbitrary constants, is completely integrable.
If $c_2=a_1=0,$ then the Hamiltonian just 
reduces to the famous Kowalewski Hamiltonian. The case 
$c_2=0$ corresponds to the Kowalewski Hamiltonian with the additional gyrostatic term. 
If $a_1=c_1$, we get the Hamiltonian function for the integrable case for the Kirchhoff equations  
found in \cite{sok1}. 

It turns out that there exists a polynomial of fourth degree, which commutes with (\ref{tailHAM}) with 
respect to the Lie-Poisson bracket 
\begin{equation}
\bigl\{ J_i\,,J_j\,\bigr\}= \varepsilon_{ijk}\,J_k\,, \qquad
\bigl\{ J_i\,,x_j\,\bigr\}= \varepsilon_{ijk}\,x_k\,,\qquad
\bigl\{ x_i\,,x_j\,\bigr\}= 0\,,\qquad i,j,k=1,2,3, \label{e3}
\end{equation}
where $\varepsilon_{ijk}$ is the standard totally skew-symmetric
tensor. These brackets 
possess two Casimir elements
\bq
I_1=(x,x), \qquad I_2=(J,x)\,, \label{caz}
\eq
where $J=(J_1, J_2, J_3)$, $x=(x_1, x_2, x_3)$ and  $(x,y)$ stands for 
the scalar product in ${\mathbb R}^3$. Since generic symplectic leaves specified by the values 
of these Casimir elements have dimension four, we need only one additional first integral for 
the Liouville integrability of the corresponding equations of motion given by the standard formulae 
\bq
J_t=\bigl\{ J\,,H\,\bigr\}, \qquad x_t=\bigl\{ x\,,H\,\bigr\}\, \label{evol}.
\eq

In this paper we present a $5\times 5$-matrix Lax pair for (\ref{evol}), (\ref{tailHAM}), which generalizes 
the corresponding Lax pair from the paper \cite{brs89}. Following \cite{mark01}, 
we find a deformation of the famous Kowalewski curve 
with respect to the additional parameter $c_2$. 

We also find generalizations of the two-field gyrostat and the $so(p,q)$-model  from \cite{brs89} and present 
Lax pairs for them.

Moreover, we find a similar $3\times 3$-matrix Lax pair for a generalized 
Goryachev-Chaplygin top whose Hamiltonian function has the form 
\bq
 \label{gorchap}
H=J_1^2+J_2^2+4 J_3^2-2 a_1 J_3-4 c_1 x_2-4 a_1 c_2 x_1+8 c_2 J_3 x_1-4 c_2^2 x_3^2.
\eq
If $c_2=0$ this Hamiltonian coincides with the usual Goryachev-Chaplygin gyrostat and our Lax pair 
reduces to the Lax representation from \cite{bk88}.
Like the Goryachev-Chaplygin gyrostat the generalization is an integrable system on the level 
$I_2=0$ only.
 In the case $a_1=c_1=0$ we get a new partially integrable (i.e. integrable on a 
special level of one of the integrals of motion) case for the Kirchhoff equations.

\section{Generalized Kowalewski top}

The Kowalewski gyrostat is defined by the Hamiltonian (\ref{tailHAM})
 with $c_2=0$.
In the paper \cite{brs89} a Lax representation \[\dfrac{d}{dt} L_{kow}=[M_{kow},\, L_{kow}]\] for this 
system has been found. The corresponding Lax matrices $L_{kow}$ and $M_{kow}$ are given by 
\bq
L_{kow}(\lambda)=\lambda\, A+B+c_1\lambda^{-1}\, C\,,\qquad M_{kow}(\lambda)=-2 \lambda\, A+ D
\eq
 where
\[A=\left(\begin{array}{ccccc} 0& 0& 0& 1& 0\\
0& 0& 0& 0&1\\
 0&0& 0&0&0\\
1&0& 0& 0& 0\\
0&1&0& 0& 0
\end{array}\right)\,,\qquad B=\left(\begin{array}{ccccc} 0& J_3& -J_2& 0 & 0\\
-J_3& 0& J_1& 0&0\\
 J_2& -J_1& 0&0&0\\
0&0& 0& 0& -J_3-a_1\\
0&0&0& J_3+a_1& 0
\end{array}\right)
\]
and
\[C=\left(\begin{array}{ccccc}
0&0&0&0&x_1\\
0&0&0&0&x_2\\
0&0&0&0&x_3\\
0&0&0&0&0\\
x_1&x_2&x_3&0&0
\end{array}\right), \qquad D=\left(\begin{array}{ccccc} 0& -4J_3-2 a_1& 2 J_2& 0& 0\\
4 J_3+2 a_1& 0& -2 J_1& 0&0\\
 -2 J_2& 2 J_1& 0&0&0\\
0&0& 0& 0&0\\
0&0&0& 0& 0
\end{array}\right).\nn\\
\]
The characteristic curve $Det(L_{kow}(\lambda)-\mu\cdot Id)=0,$ where $Id=diag(1,1,1,1,1)$ is the 
unit matrix provides a 
complete set of first integrals for the Kowalewski gyrostat \cite{brs89}.  

In this paper we consider a matrix of the following form 
\bq
L(\lambda,\mu)=L_1(\lambda)+\mu\cdot L_2(\lambda), \, \label{LAX}
\eq
where \[L_1(\lambda)=L_{kow}(\lambda)+c_2 X, 
\qquad L_2(\lambda)=-Id+c_2 \lambda^{-1} Y,\]
and
\[X=\left(\begin{array}{ccccc}
0&0&0&0&0\\
0&0&0&0&0\\
0&0&0&0&0\\
0&0&0&0&-x_1\\
0&0&0&x_1&0
\end{array}\right), \qquad Y=\left(\begin{array}{ccccc} 
0& 0& 0& 0&x_1\\
0&0& 0& 0&x_2\\
0&0&0& 0&x_3\\
0&0& 0& 0&0\\
-x_1&-x_2&-x_3& 0& 0
\end{array}\right).\nn\\
\]
Obviously, if $c_2=0$ then this matrix $L(\lambda,\mu)$ coincides with $L_{kow}(\lambda)-\mu\cdot Id.$

It is easy to verify that for operator ({\ref{LAX}}) the following symmetry properties hold:
\bq \label{sym}
L(\lambda,\mu)=-L^T(-\lambda,-\mu), \qquad 
L(\lambda,\mu)=V^{-1}L(-\lambda, \mu)V, 
\eq
where $V=diag(1,1,1,-1,-1).$

A simple calculation shows that the algebraic curve ${\cal C}$: \, $Det\Big(L(\lambda,\mu)\Big)=0$ can be 
written in the following form 
\begin{equation}
\begin{array}{c}
{\cal C}: \quad d_2(\lambda^2)\, \mu^4+d_1(\lambda^2)\, \mu^2+d_0(\lambda^2)=0, \\[5mm]
d_2=\lambda^2+c_2^2 I_1, \qquad 
d_0=\lambda^6-H \lambda^4+I_4 \lambda^2-c_1^2 I_2^2, \\[4mm]
d_1=-2 \lambda^4+(H+a_1^2-c_2^2 I_1)\lambda^2+(c_2^2I_2^2-c_1^2 I_1), 
\label{curve}
\end{array}
\end{equation}
where 
\bq
H= J_1^2+J_2^2+2 J_3^2-2 c_1 x_2+2 a_1 J_3
+2 c_2 (J_3 x_1-x_3 J_1)
\label{nham}
\eq
and 
\ben
I_4&=& x_1\Bigl(x_1(J_2^2+J_3^2-J_1^2)-2(x_3J_3+x_2J_2)J_1\Bigr)c_2^2
+(x_2^2+x_3^2)c_1^2\label{sint}\\
\nn\\
&+&2\Bigl(c_1x_1(x_3J_2-x_2J_3)+(J_3+a_1)(x_1(J_2^2+J_3^2)
-(x_2J_2+x_3J_3)J_1)\Bigr)c_2\nn\\
\nn\\
&-&2\Bigl(x_2(J_2^2+J_3^2+a_1J_3)+(x_1J_1-a_1x_3)J_2\Bigr)c_1
+(J_1^2+J_2^2+J_3^2)(J_3+a_1)^2\nn
.
\en

The following statement can be proved by a straightforward calculation. 
\begin{prop}
 \quad 
$\{ H,\, I_4\}=0.$
\end{prop}

It follows from Proposition 1 that the functions $I_3=H$ and $I_4$ are integrals 
of motion in involution and the corresponding Hamiltonian system is completely
integrable. Notice that the Hamiltonian (\ref{nham}) up to a canonical transformation of the form
$$
J_1\rightarrow J_1+c_2 x_3, \qquad J_2\rightarrow J_2, \qquad J_3\rightarrow J_3-c_2 x_1,
$$ 
coincides with (\ref{tailHAM}).
 
The next theorem describes Lax structures related to the operator (\ref{LAX}).
\begin{thrm}
The flow with the Hamiltonian (\ref{nham}) is equivalent to the
following matrix differential equations
\bq
\frac{d}{dt}L_i=L_i\,M(\lambda)+M^T(-\lambda)\, L_i, \qquad i=1,2,
\label{laxtr}
\eq
where
 \bq M=M_{kow}+W,\qquad
W=2c_2\left( \begin{array}{ccccc}
  0& x_1& 0& 0& 0\\
  0& x_2& 0& 0& 0\\
  0& x_3& 0& 0& 0\\
  0& 0& 0& 0& -x_1\\
  0& 0& 0& 0& -x_2
\end{array}
\right) \,,\qquad
\eq
and the superscript $T$ stands for matrix transposition.
\end{thrm}

The relations (\ref{laxtr}) imply  that the operators 
\bq \label{newlax2}
L_+=L_1(\lambda)\,L_2^{-1}(\lambda)
, \qquad 
L_-=L_2^{-1}(\lambda)\,L_1(\lambda)
\eq
satisfy the usual Lax equations
\bq 
\frac{d}{dt} L_+=\left[
L_+,\, -M^T(-\lambda)\right], \quad \frac{d}{dt} L_-=\left[
L_-,\, M(\lambda) \right].  \label{2flax}
\eq
An explicit formula for $L_2^{-1}$ can be written as follows:
$$
L_2^{-1}=-\Big(Id+\frac{1}{\lambda}Y+\frac{1}{\lambda^2}Y^2+\frac{1}{\lambda^3 \Delta}Y^3
+\frac{1}{\lambda^4 \Delta}Y^4\Big),
$$
where 
\[\displaystyle \Delta=-Det(L_2)=1+\frac{c_2^2 I_1}{\lambda^2}\]
is a Casimir function.

It is clear that the determinant curves $Det(L(\lambda,\mu))=0,$
$Det(L_+(\lambda)-\mu Id)=0,$ and 
$Det(L_-(\lambda)-\mu Id)=0 $ 
coincide with each other up to inessential multipliers. 
Therefore one can use one of the operators $L_{\pm}$ 
to generalize the results of \cite{brs89},
though to our taste the operator $L$ looks more 
elegant than the operators $L_{\pm}.$

{\bf Remark.} The Lax triads (\ref{laxtr}) and the Lax matrices of the form 
$L_1(\lambda)L_2^{-1}(\lambda)$ have arisen in the case of the 
relativistic Toda lattice (see \cite{sur93} and references within). However 
in that case the operator $L_1$ was the same for the initial 
and deformed models whereas in our case $L_1$ must be deformed also.

According to \cite{brs89} let us consider the projection of curve 
$C$ (\ref{curve}) given by the change of variables $z=\lambda^2$
\[
{\cal C}_1:\qquad d_2(z)\mu^4+d_1(z)\mu^2+d_0(z)=0\,.
\]
The genus of ${\cal C}_1$ reduces from $3$ to $2$ in the following 
two cases:  $a_1=I_2=0$ or $a_1=c_1=0$. The latter case corresponds to the new integrable case for the Kirchhoff 
equations found in \cite{sok1}.

Let us consider the first case. Following \cite{mark01}, we can easily find a deformation of 
the famous Kowalewski curve with respect to the parameter $c_2$. Namely, the following transformation 
\bq
\mu=\frac{y}{x^2+(H+I_1 c_2^2)\,x+I_4-I_1 c_1^2}, \qquad 
z=\frac{I_1 x (c_1^2-c_2^2 x)}{x^2+(H+I_1 c_2^2)\,x+I_4-I_1 c_1^2}
\eq
brings ${\cal C}_1$ to the normal form
\bq
{\cal C}_2: \qquad
y^2=x\left(x^2+Hx+I_4\right)\left(x^2+(H+c_2^2 I_1)x+I_4-c_1^2 I_1\right)\,.
\label{hcurve1}
\eq
This curve differs from the corresponding curve from \cite{brs89} 
by the third factor only. According to
\cite{mark01} we can use Richelot's transformation of the curve ${\cal
C}_2$ in order to get another hyperelliptic curve
\bq
\widetilde{\cal C}_2:\qquad \eta^2 =
\Bigl(c_2^2 \zeta^2-2 c_1^2 \zeta-H c_1^2-I_4 c_2^2\Bigr)
\Bigl(\zeta^2-I_4+I_1 c_1^2\Bigr)\Bigl(\zeta^2-I_4\Bigr)
\label{kowcurve}
\eq
which is a deformation of the usual Kowalewski curve. As above, the genus of
this curve is not changed and there is a difference in the first
factor only. Of course, if $c_2=0$ and $c_1=1$ then the curve
$\widetilde{\cal C}_2$ coincides with the Kowalewski curve \cite{mark01}.

In the second case the normal form of ${\cal C}_1$ is 
\bq
{\cal C}_3: \qquad
y^2=x\left(x^2+Hx+I_4\right)\left(x^2+(H+c_2^2 I_1)x+I_4+c_2^2 I_2^2\right)\,
 .
\label{hcurve2}
\eq
and Richelot's transformation gives rise to
\bq
\widetilde{\cal C}_3:\qquad \eta^2 =
\Bigl(I_1 \zeta^2+2 I_2^2 \zeta+H I_2^2-I_4 I_1\Bigr)
\Bigl(\zeta^2-I_4-I_2^2 c_2^2\Bigr)\Bigl(\zeta^2-I_4\Bigr)
\label{kowcurve1}
\eq
It is interesting to note a duality between $I_2$ and $c_1$ 
in Case 1 and Case 2. By analogy with the Kowalewski case one can expect that (\ref{kowcurve}) 
and (\ref{kowcurve1}) are separation curves for the corresponding cases.

\section{Generalized two-field gyrostat}

In the two-field case we have three vectors $J=(J_1, J_2, J_3)$, $x=(x_1, x_2, x_3),$ and  
$y=(y_1, y_2, y_3).$
The Lie-Poisson bracket
 is given by
\begin{equation}
\begin{array}{l}
\bigl\{ J_i\,,J_j\,\bigr\}= \varepsilon_{ijk}\,J_k\,, \qquad
\bigl\{ J_i\,,x_j\,\bigr\}= \varepsilon_{ijk}\,x_k\,,\qquad
\bigl\{ x_i\,,x_j\,\bigr\}= 0\,\\[4mm]
\bigl\{ J_i\,,y_j\,\bigr\}= \varepsilon_{ijk}\,y_k\,,\qquad
\bigl\{ y_i\,,y_j\,\bigr\}= 0,\qquad
\bigl\{ x_i\,,y_j\,\bigr\}= 0,\qquad i,j,k=1,2,3. \label{twoe3}
\end{array}
\end{equation}
The Casimir functions are $(x,\,x), \,(x,\,y),$ and $(y,\,y)$.

We claim that the Hamiltonian function 
\bq
 \label{twoHAM}
H= J_1^2+J_2^2+2J_3^2-2 c_1 x_2-2 b_1 y_1+2 a_1 J_3+2 c_2 (J_3 x_1- x_3 J_1)-2 b_2 (J_3 y_2-J_2 y_3)\,
\eq
gives rise to a completely integrable model if 
$$
c_1 b_2-b_1 c_2=0.
$$
Although the parameters can be normalized by scalings, we prefer to keep them because 
all reductions and limits are more obvious in this form. 
In the case $b_1=b_2=0$ the Hamiltonian function (\ref{twoHAM}) coincides with (\ref{nham}). 
If $a_1=c_1=b_1=0$ we have a new homogeneous quadratic integrable Hamiltonian. 

Two necessary additional integrals of motion are the coefficients at $\lambda^4$ and $\lambda^2$ of 
the algebraic curve $Det\Big(L(\lambda,\mu)\Big)=0$, where $L=L_1+ \mu L_2$ ,
$$
L_1=\lambda A+B+\widehat{X}+\lambda^{-1} \widehat{C} , 
\qquad L_2=-Id + \lambda^{-1} \widehat{Y}.
$$
The matrices $A$ and $B$ are defined in the previous section and
\[
\widehat{X}=\left(\begin{array}{ccccc} 0& 0& 0& 0 & 0\\
0& 0& 0& 0&0\\
 0&0& 0&0&0\\
0&0& 0& 0& -c_2 x_1+b_2 y_2\\
0&0&0& c_2 x_1-b_2 y_2& 0
\end{array}\right)
,
\]
\[\widehat{C}=\left(\begin{array}{ccccc}
0&0&0&b_1 y_1&c_1 x_1\\
0&0&0&b_1 y_2&c_1x_2\\
0&0&0&b_1 y_3&c_1x_3\\
b_1 y_1&b_1 y_2&b_1 y_3&0&0\\
c_1 x_1&c_1 x_2&c_1 x_3&0&0
\end{array}\right), \qquad 
\widehat{Y}=\left(\begin{array}{ccccc} 
0& 0& 0& b_2 y_1&c_2 x_1\\
0&0& 0& b_2 y_2&c_2 x_2\\
0&0&0& b_2 y_3&c_2 x_3\\
-b_2 y_1&-b_2 y_2&-b_2 y_3& 0&0\\
-c_2 x_1&-c_2 x_2&-c_2 x_3& 0& 0
\end{array}\right).
\]
The operators $L_1$ and $L_2$ satisfy (\ref{sym}) and (\ref{laxtr}),
with
 \bq M=M_{kow}+\widehat W,\qquad
\widehat{W}=2\left( \begin{array}{ccccc}
  b_2 y_1&c_2 x_1& 0& 0& 0\\
  b_2 y_2&c_2 x_2& 0& 0& 0\\
  b_2 y_3&c_2 x_3& 0& 0& 0\\
  0& 0& 0& -b_2 y_1& -c_2 x_1\\
  0& 0& 0& -b_2 y_2& -c_2 x_2
\end{array}
\right) \,.\nn
\eq

\section{Generalized $q$-field top}
In this Section we present an integrable polynomial deformation of the $so(p,q)$ Kowalewski system. 
If $p=3$ and $q=2,$ then the deformed Hamiltonian coincides with (\ref{nham}), where 
$a_1=0, \, c_1=c_2=1$.

Recall that for any $p$ and $q$ such that $q\leq p$ the Hamiltonian
\[H_{old}=\dfrac12\Bigl(\sum_{i,j=1}^p l_{ij}^2+\sum_{i,j=1}^q l_{ij}^2\Bigr)-2\sum_{i=1}^q F_{ii}\,.\]
defines the so called $so(p,q)$- analog of the Kowalewski top. Here $c$ is arbitrary constant, 
and dynamical variables $l_{ij}$ and $F_{ij}$ are entries of 
a skew-symmetric $p\times p$ matrix $l$ and a $p\times q$ matrix $F$.

The $(p+q)\times(p+q)$ matrix Lax pair for this system found in \cite{brs89} is given by 
\[
L_{old}(\lambda)=\lambda\, A+B+\lambda^{-1}\, C\,,
\qquad M_{old}(\lambda)=-2 \lambda\, A+ D
\]
where
\[A=\left( \begin{array}{cc}0&E\\E^T&0\end{array}
\right)\,,\qquad
B=\left( \begin{array}{cc}-l&0\\0&E^T l E\end{array}
\right)\,,\qquad
C=\left( \begin{array}{cc}0&F\\F^T&0\end{array}
\right)
\,,\qquad
D=\left( \begin{array}{cc}\omega&0\\0&0\end{array}
\right)\,.\]
Here $E$ is a $p\times q$ matrix, whose non-zero entries are $E_{ii}=1, \, i\leq q.$
The entries of the matrix $\omega$ are defined as follows: 
$\omega_{ij}=4l_{ij}$, if $i,j\leq q$ and otherwise
$\omega_{i,j}=2l_{ij}$.

Let us consider a deformation of this Hamiltonian
\[
H=H_{old}-2\sum_{i=1}^p \sum_{j=1}^q F_{ij}l_{ij}\,.
\]
The corresponding Hamiltonian flow is equivalent to the matrix
differential equations (\ref{laxtr}) with
matrices 
\[
L(\lambda,\mu)=L_1(\lambda)+\mu L_2(\lambda)\,,
\qquad M(\lambda)=M_{old}(\lambda)+W\,,
\]
where 
\[L_1(\lambda)=L_{old}+X\,,\qquad
  L_2(\lambda)=-Id+\lambda^{-1}Y\]
and
\[
Y=\left( \begin{array}{cc}0&F\\-F^T&0\end{array}
\right) \,, \quad
X=\dfrac12\Bigl((C-Y)A-A(C+Y)\Bigr)\,,\qquad
W=\Bigl((C+Y)A-A(C+Y)\Bigr)\,.\]

\section{Generalized Goryachev-Chaplygin top}
In this section we establish analogous structures for the simpler case of the 
Goryachev-Chaplygin top. 

Let us consider the following $3\times 3$ matrix 
$L(\lambda,\mu)=L_1(\lambda)+\mu L_2(\lambda),$ where
\bq
L_1=\lambda S+J+c_2 B+ic_1\lambda^{-1}\,X, \qquad L_2=Id-c_2 \lambda^{-1}\,Y,
\eq
and
\ben
&S=\left(
\begin{array}{ccc}
 0& 0& 0\\ 0& 0& -2\\
 0& 2&0\end{array}\right)\, , \qquad 
 &J=\left(
\begin{array}{ccc}
 0& 0& -J_1-iJ_2\\ 0& -2J_3+a_1& 0\\
 -J_1+iJ_2& 0&2J_3\end{array}\right)\,
\nn\\
\nn\\ 
&X=\left(
\begin{array}{ccc}0& x_3& 0\\
x_3& 0& x_1+ix_2\\
0& x_1-ix_2& 0
\end{array}\right)
, \qquad &Y=\left(
\begin{array}{ccc}0& x_3& 0\\
-x_3& 0& -x_1-ix_2\\
0& x_1-ix_2& 0
\end{array}\right)
,
\nn
\en
$$
B=\dfrac12\Bigl((X-Y)S-S(X+Y)\Bigr)
=4x_1\left(
\begin{array}{ccc}
 0& 0& 0\\ 0& 1& 0\\
 0& 0&0\end{array}\right).
 $$
Everywhere we assume that $I_2=0$. If $c_2=0,$ this $L$-operator coincides with the operator found 
in \cite{bk88}. 

The corresponding spectral curve is
\[
(\lambda^2+c_2^2 I_1)\, \mu^3-a_1 \lambda^2 \mu^2+(4 \lambda^4-H \lambda^2+c_1^2 I_1) \mu-I_4 \lambda^2=0
\]
where
\ben
H&=&J_1^2+J_2^2+4 J_3^2-2 a_1 J_3-4 c_1 x_2 +4 c_2 (J_1 x_3-2 J_3 x_1)\,,\label{nham1}\\
\nn\\
I_4&=&(J_1^2+J_2^2)(2 J_3-4 c_2 x_1-a_1)+4 c_1 J_2 x_3\,.
\en

It is easy to verify that, for the operator $L,$ the following symmetry properties hold:
\bq \label{sym1}
L^*(-\lambda, \mu)=L(\lambda,\mu), \qquad L(-\lambda, \mu)=V L(\lambda,\mu)V^{-1}, 
\eq
where $V=diag(1,-1,1)$ and $*$ means Hermitian conjugation.

\begin{thrm}
The flow with the Hamiltonian (\ref{nham1}) is equivalent to the
following matrix differential equations
\bq
\frac{d}{dt}L_i=L_i\,M(\lambda)+M^*(-\lambda)\,L_i, \qquad i=1,2,
\label{laxtr1}
\eq
where 
$M=2 i (\lambda S+W),$
$$
W=\left(
\begin{array}{ccc}
 -J_3& 0& -J_1-iJ_2\\ 0& 0& 0\\
 -J_1+iJ_2& 0&4 J_3-a_1\end{array}\right)-2c_2
 \left(
\begin{array}{ccc}
 0& 0& x_3\\ 0& ix_2& 0\\
 0& 0&2 x_1-ix_2\end{array}\right)\,.
$$
\end{thrm}

After the canonical transformation  
$$
J_1\rightarrow J_1-2 c_2 x_3, \qquad J_2\rightarrow J_2, \qquad J_3\rightarrow J_3+2 c_2 x_1,
$$ 
this Hamilton function takes the form
 (\ref{gorchap}).
\vskip1truecm

{\bf Acknowledgments.} The authors are grateful to the Newton Institute (Univ. of Cambridge) 
for its hospitality. The
 research was partially supported by RFBR grants 99-01-00294 and 99-01-00698,
INTAS grants 99-1782 and 99-01459, and EPSRC grant GR K99015.


\begin{thebibliography}{10}

\bibitem{bk88}
A.I. Bobenko and V.B. Kuznetsov,
\newblock{Lax representation and new formulae for the Goryachev-Chaplygin top},
\newblock{\em J.Phys.A.}, v. 21, p. 1999, 1988.

\bibitem{brs89}
A.G. Reyman and M.A. Semenov-{T}ian {S}hansky,
\newblock{Lax representation with a spectral parameter for the Kowalewski top
and its generalizations},
\newblock{\em Lett.Math.Phys}, v.14, p.55, 1987. 

A.I. Bobenko and A.G. Reyman and M.A. Semenov-{T}ian {S}hansky,
\newblock{The Kowalewski top 99 years later: a Lax pair, generalizations and explicit solutions},
\newblock{\em Commun.\-Math.\-Phys.}, v.122, p.321, 1989.

\bibitem{sur93}
Yu. B. Suris,
\newblock{On the bi-Hamiltonian structure of Toda and relativistic Toda lattices},
\newblock{\em Phys. Lett. A.}, v.180, p.419, 1993.


\bibitem{mark01}
D. Markushevich,
\newblock{Kowalewski top and genus-2 curves},
\newblock
{\em J.\-Phys.A.}, v.34, p.2125, 2001.

\bibitem{sok1}
V.V. Sokolov,
\newblock{A new integrable case for the 
Kirchhoff equation},
\newblock{\em Teor.Math.Phys.}, v.128(2), p.31, 2001.

\bibitem{sok2}
V.V. Sokolov,
\newblock{A generalized Kowalevski Hamiltonian and new integrable 
cases on
 $e(3)$ and $so(4)$},
\newblock{ Preprint \em nlin.SI/0110022},  2001.


\end{thebibliography}
\end{document}